\documentstyle[12pt]{article}
\def\beq{\begin{equation}}
\def\eeq{\end{equation}}

\begin{document}

\begin{titlepage}
\begin{flushright}
BA-98-32 \\
INFN-FE 12-98 \\ 
July 26, 1998 \\
hep-ph/9807502 
\end{flushright}

\begin{center}
{\Large\bf Phenomenology with Supersymmetric \\

Flipped SU(6)
}
\end{center}
\vspace{0.5cm}
\begin{center}
{\large Qaisar Shafi$^{a}$\footnote {E-mail address:
shafi@bartol.udel.edu} {}~and
{}~Zurab Tavartkiladze$^{b, c}$\footnote {E-mail address:
tavzur@axpfe1.fe.infn.it} }
\vspace{0.5cm}

$^a${\em Bartol Research Institute, University of Delaware,
Newark, DE 19716, USA \\

$^b$Istituto Nazionale di Fisica Nucleare, 
Sezione di Ferrara, 44100 Ferrara, Italy  \\

$^c$ Institute of Physics, Georgian Academy of Sciences,
380077 Tbilisi, Georgia}\\
\end{center}

\vspace{0.5cm}

\begin{abstract} 

The supersymmetric flipped $SU(6)\times U(1)$ gauge symmetry can arise
through compactification of the ten dimensional $E_8\times E_8$
superstring theory. We show how realistic phenomenology can emerge from
this theory by supplementing it with the symmetry
${\cal R}\times {\cal U}(1)$, where ${\cal R}$ denotes a discrete 
`R'-symmetry. The well-known doublet-triplet splitting problem is resolved
to `all orders' via the pseudo-Goldstone mechanism, and the GUT scale
arises from an interplay of the Planck and supersymmetry breaking scales.

The symmetry ${\cal R}\times {\cal U}(1)$ is also important for
understanding the fermion mass hierarchies as well as the magnitudes of
the CKM matrix elements. Furthermore, the well known MSSM parameter 
$\tan \beta $ is estimated to be of order unity, while the proton
lifetime ($\tau_p\sim 10^2\tau_{pSU(5)}$) is consistent with observations.
Depending on some parameters, $p\to K\mu^+ $ can be the dominant decay
mode.

Finally, the observed solar and atmospheric neutrino `anomalies' require
us to introduce a `sterile' neutrino state. Remarkably, the 
${\cal R}\times {\cal U}(1)$ symmetry protects it from becoming heavy,
so that maximal angle $\nu_{\mu }$ oscillations into a sterile state can
explain the
atmospheric anomaly, while the solar neutrino puzzle is resolved via the
small angle $\nu_e-\nu_{\tau }$ MSW oscillations.
The existence of some ($\sim 15$-$20\% $ of critical energy density)
neutrino hot dark matter is also predicted.

\vspace{0.3cm}

\hspace{-0.5cm} PACS:~$12$.$60$. Jv; $12.10.$ Dm; 12.15. Ff; 14.60. Pq;
14.60. St
 
\end{abstract}

\end{titlepage}
\newpage
\section{Introduction}
 It is a curious fact that the most well known grand unified theories
(GUTs)
$SU(5)$ and $SO(10)$ do not readily arise within the framework of the
simplest
superstring theories. On the contrary, compactification of the ten
dimensional $E_8 \times E_8$ heterotic superstring theory say on a Calabi-
Yau
manifold leads one to a variety of subgroups of $E_6$, such as $SU(3)^3$
and $SU(6) \times U(1)$ (or flipped $SU(6)$). The group $SU(3)^3$ as a
grand unified
symmetry has attracted \cite{gia1, laz} a certain amount of attention, but the
flipped
$SU(6)$ case has so far been more or less ignored \cite{str} . In this paper we
hope to
remedy this situation by discussing how a realistic `low energy' phenomenology
can emerge from flipped $SU(6)$.
 
The fact that $SU(5)$ and $SO(10)$ do not readily appear from
superstrings may
be a blessing in disguise. Consider, for instance, the 
doublet-triplet (DT)
splitting problem. A number of mechanisms for resolving this thorny problem
have been proposed. These include the missing partner \cite{dim, flip} 
and the missing
VEV \cite{dim1}
mechanisms. However, their implementation results in a grand unified framework
which is far from `simple'. A much more attractive possibility of realizing
a pair of light electroweak doublets is provided by the pseudo-Goldstone
mechanism . Here the light doublets emerge as the pseudo-Goldstone modes 
from an `accidental' and larger global symmetry of the Higgs superpotential.
It turns out that this idea is hard (if not impossible) 
to realize in $SU(5)$
and $SO(10)$, but is readily implemented in GUTs such as $SU(6)$ 
\cite{ino, golds} and 
$SU(3)^ 3$ \cite{gia1}.

 In this paper we will see that the pseudo-Goldstone
 mechanism for resolving the
DT splitting problem can be neatly realized within the framework of flipped
$SU(6)$. Moreover, it also becomes possible to understand how the GUT
scale
can emerge from an interplay of the Planck and supersymmetry 
breaking ($\sim $~TeV)
scales. We also study fermion masses and mixings in this scheme. It turns out
that the well known $SU(5)$ relation $m_b = m_{\tau }$ holds in the
flipped $SU(6)$ 
scheme presented here. An important role is played 
by the symmetry ${\cal R} \times {\cal U}(1)$
that we impose in addition. This symmetry, among other things, helps implement
the pseudo-Goldstone mechanism and distinguishes the families so that the
fermion masses and mixings (especially the observed hierarchies) can be
explained.

 The flipped $SU(6)$ scheme has a number of testable predictions. The
well-known
MSSM parameter $\tan \beta $ is estimated to be of order unity. One expects the
dominant
proton decay mode to be $p\to K^0\mu^+$ , with a rate that is suppressed
relative
to the dominant $SU(5)$ mode ($p\to K^+ \nu_{\mu }$) by about two orders of
magnitude. It
is
worth noting that the symmetry ${\cal R} \times {\cal U}(1)$ plays an
important role in the
suppression of all dimension five Planck scale induced operators.

 Finally, the neutrino sector of flipped $SU(6)$ turns out to be quite  
interesting  and unique. It turns out that in order to explain the recent
Superkamiokande results on atmospheric neutrinos 
\cite{sup} as well as the solar
neutrino puzzle, one is led to introduce one sterile neutrino state which
is kept `light' thanks to the presence of the 
${\cal R} \times {\cal U}(1)$ symmetry! One finds
that the atmospheric neutrino anomaly is explained via  
$\nu_{\mu }$-$\nu_s$ 
oscillations with maximal mixing, while the resolution of the 
solar neutrino puzzle relies on the
small angle MSW oscillations of $\nu_e$ into $\nu_{\tau }$ \cite{msw}.
It is worth emphasizing  that this scheme implies the
existence of some hot dark matter
($\sim 15$-$20\% $ of critical energy density).

 The paper is organized as follows: In section 2 we describe the salient 
features of flipped $SU(6)$, the symmetry breaking pattern, and details of
how
the pseudo-Goldstone mechanism is realized in this scheme. Section 3 is devoted
to understanding the fermion masses and mixings, especially the hierarchies.
In section 4 we discuss proton decay including suppression of Planck scale
induced dimension five operators. The discussion about neutrinos is contained
in section 5 and the conclusions are presented in section 6.


\section{Flipped $SU(6)\times U(1)$ Model}

  The flipped $SU(6)$ models are perhaps best motivated from the 
compactification of the ten dimensional heterotic 
$E_8\times E_8$ superstring
theory \cite{str}, \cite{can}-\cite{str1},  on a suitable 
Calabi-Yau manifold $K$ . The compactification
process leads to a four dimensional theory with 
$E_6$ gauge symmetry and $N=1$ 
supersymmetry, a certain number of left-handed superfields belonging to
the $27$ and $\overline {27}$ 
representations of $E_6$, and a group of discrete symmetries
(the isometries of $K$) \cite{can}.  With a non-simply 
connected $K$ it is possible induce
the breaking of $E_6$ with Wilson loops to a subgroup $H$. 
It turns out that the
simplest constructions lead to $H=SU(3)^3$ or 
$H=SU(6)\times U(1)$ \cite{laz}.

It is perhaps fair to state that so far there does not exist a single example
of a string derived four dimensional theory which provides a satisfactory
explanation of the most important issues in phenomenology . These include
the gauge hierarchy problem, proton stability, fermion mass hierarchies and
mixings, etc. A route chosen by many is the so-called `string inspired'
approach, in which the choice of the underlying gauge symmetry is dictated
by some string theory, which is then supplemented by additional symmetries
so that a realistic `low energy ' scenario can be realized, hopefully with
some predictions that can be experimentally tested. Our approach here follows
this philosophy and is similar to the one pursued in 
the $SU(3)^3$ case \cite{gia1, laz}.
For earlier works on flipped $SU(6)$ see refs. \cite{str}, \cite{str1} , 
while $SU(6)\times  SU(2)_{L,R}$ has been
discussed in ref \cite{haba}.

\subsection{ Pseudogoldstone Mechanism and \\
 GUT
Scale in Flipped $SU(6)$}


 Under the $SU(6)\times U(1)$ gauge symmetry, the chiral $27$-plet 
of $E_6$ transforms
as  $15_0+\bar 6_1+\bar 6_{-1}$, where the subscripts 
refer to the $U(1)$ charge.
More explicitly,
               
$$
15_0=(q, d^c, \nu^c, g, L )_0~,
$$
$$
\bar 6_1=(u^c, l, N)_1~,
$$
\begin{equation}  
\bar 6_{-1}=(\bar g, \bar L, e^c)_{-1}~.
\label{rep}
\end{equation} 
Decomposition under the $SU(5)\times U(1)'$ yields:

$$
15_0=10_1+5_{-2}~,~~~~~~\bar 6_1=\bar 5_{-3}+1_0~,
$$
\begin{equation}  
\bar 6_{-1}=\bar 5_2+1_5~,
\label{rep1}
\end{equation}
which is just the chiral content of flipped $SU(5)$ \cite{flip}, 
supplemented by the
singlet $1_0$ and the pair $\bar 5_2+5_{-2}$ of vector states. We will focus
in this section on the Higgs sector of the theory , the implementation
of the pseudo-Goldstone mechanism through the introduction of an
additional
symmetry ${\cal R}\times {\cal U}(1)$, and also show how the GUT scale can arise
from an interplay
of $M_P$ and $m_{3/2}$, the two basic scales in the theory.

For the breaking of  
$SU(6)\times U(1)$ to  $G_{SM}$ 
($\equiv SU(3)_C\times SU(2)_W\times U(1)_Y$)
it is enough to introduce 
the following Higgs supermultiplets:
 
$$
\overline {\Psi } \sim \overline {15}_0~,~~~~~~\Psi \sim 15_0~, 
$$
\begin{equation}
\bar H \sim \bar 6_{1}~,~~~~~~
H \sim 6_{-1}~.
\label{screp}
\end{equation}  
The $\overline {\Psi }+\Psi$ and $\bar H +H$ contain the same fragments
as supermultiplets of the chiral matter of $15_0$ and $\bar 6_1$ 
(see (\ref{rep})). 
The VEVs of the fragments  $\overline {\nu }^c_{\overline \Psi }$ and 
${\nu }^c_{\Psi }$ from $\overline \Psi $ and $\Psi $   
break  $SU(6)\times 
U(1)$ to $SU(4)\times SU(2)_W\times U(1)_1$, while
the VEVs of the fragments $\bar N_H$, $N_{\bar H}$ from $H$, $\bar H$
break $SU(4)\times SU(2)_W\times U(1)_1$ down to $G_{SM}$.
The VEVs in the group space have the following directions:
$$
\langle \overline {\Psi}^{mn}\rangle =
\langle \Psi_{mn} \rangle =
\frac{V}{\sqrt 2}\left (\delta_{4m}\delta_{5n}-
\delta_{4n}\delta_{5m} \right )~,
$$
\begin{equation}
\langle \bar H^m \rangle =
\langle H_m \rangle =v\delta_{m6}~,
\label{VEVs}
\end{equation} 
where $m, n$ are $SU(6)$ indices, the indices $4, 5$ correspond 
to the $SU(2)_W$
group, and the index $6$ is a broken degree of freedom of $SU(6)$.

Due to the $SU(6)\times U(1)$,   
at the renormalizable
level there are no couplings between the $\Psi+\overline {\Psi}$ and 
$\bar H+H$ superfields. Thus, the renormalizable superpotential has an 
$SU(6)^2\times U(1)$ global symmetry\footnote{The factor $U(1)$
appears only in the first power because the 
$\overline {\Psi}, \Psi$ superfields do not carry the $U(1)$ charges.}. 
The existence of  this global symmetry leads to the  possibility
of the realization of the pseudo-Goldstone mechanism.
Let us assume that the scalar part of the superpotential has 
$SU(6)^2\times U(1)$
global symmetry up to some desired level of the nonrenormalizable terms.
Without considering the detailed form of the scalar superpotential
and through  simple counting of the numbers of broken generators of the 
local $SU(6)\times U(1)$ group and Goldstone modes which correspond to the
breakdown of the $SU(6)^2\times U(1)$ global symmetry of the 
superpotential,
one can easily make sure that one $SU(2)_W$ doublet-antidoublet
pair emerge as a pseudo-Goldstone mode and its lightness is guaranteed
by SUSY and Goldstone theorem\footnote{Discussions 
of these and other relevant issues can be found in the original works
\cite{ino}  and in ref.  \cite{golds} as well.}.

As far as the decoupled and unphysical states are concerned the necessary
terms which must be included in the superpotential are:
\begin{equation}
W_1=\frac{\lambda_1}{12\sqrt 2}\Psi^3+
\frac{\lambda_2 }{12\sqrt 2}\overline {\Psi }^3~. 
\label{w1}
\end{equation}
Along the directions (\ref {VEVs}) this superpotential 
is flat and the values of $V$ and $v$ are not fixed.
Substituting in (\ref {w1}) the VEVs of $\overline {\Psi}+\Psi $ 
superfields the mass terms will have the form:

\begin{equation}
W_m=\lambda_1Vd^c_{\Psi }\cdot g_{\Psi}+
\lambda_2V\bar d^c_{\overline {\Psi} }\cdot \bar g_{\overline {\Psi }} . 
\label{wm}
\end{equation}
So, the triplets (antitriplets) are decoupled after symmetry breaking.
The states $\bar q_{\overline {\Psi} }+q_{\Psi }$ and 
$u^c_{\bar H}+\bar u^c_{H}$ are the  Goldstone modes. 
In addition,
one superposition of doublets (antidoublets) $L_{\Psi}$ and $l_{\bar H}$
($\bar L_{\overline {\Psi}}$ and $\bar l_{ H}$) is genuine Goldstone:

\begin{equation}
h_d^{G}=\frac{vh_{\Psi }+\sqrt 2Vh_{\bar H}}{\sqrt{v^2+2V^2}}~,~~~~~
h_u^{G}=\frac{vh_{\overline {\Psi }}+\sqrt 2Vh_H}{\sqrt{v^2+2V^2}}~,
\label{gst}
\end{equation} 
The massless pseudo-Goldstone states are given by:
\begin{equation}
h_d=\frac{vh_{\Psi }-\sqrt 2Vh_{\bar H}}{\sqrt{v^2+2V^2}}~,~~~~~
h_u=\frac{vh_{\overline {\Psi }}-\sqrt 2Vh_H}{\sqrt{v^2+2V^2}}~.
\label{dst}
\end{equation}

In order to guarantee the DT hierarchy we have to exclude the dangerous
higher order mixing terms between the $\Psi $ ($\overline {\Psi }$)
and $H$ ($\bar H$) fields which do not respect the global 
$SU(6)^2\times U(1)$ symmetry and spoil the hierarchy. This is readily 
achieved by introducing suitable discrete or continuous symmetries.
Our task then is to prescribe some transformation properties to the
scalar supermultiplets in such a way as to obtain `all order' hierarchy 
while preserving
the  terms of (\ref{w1}) . In order to suitably 
accommodate the `matter' sector
(described in the next section), we introduce two additional singlet states
$X$ and $Z$ and also ${\cal R({\cal Z})}\times {\cal U} (1)$ 
symmetry (${\cal R(Z)}$ 
is a discrete $R$ symmetry) under which the superfields transform
as:
\begin{equation}
\phi_i \to e^{{\rm i}Q_{\phi_i }}\phi_i~,
\label{sym}
\end{equation}  
where $Q_{\phi_i }$ denotes the  ${\cal R(Z)}$ and 
${\cal U}(1)$ charges of the
$\phi_i $ superfield. The transformation 
properties of the scalar superfields and superpotential 
are presented in Table 
\ref{t:scalar}. This prescription of the charges will guarantee `all
order' hierarchy, generation of scale of GUT and, as we will see
in section 3, give top Yukawa coupling of order one. We have taken
one of the simple choice, of the $Q_i$ charges of 
appropriate superfields,
which gives solution of these fundamental problems, however
other choices are also possible.

\begin{table}
\caption{${\cal R}({\cal Z})\times {\cal U}(1)$ charges of the scalar
superfields
and the superpotential.
$\alpha =\frac{2\pi }{7}$ and  $R$ 
is an undetermined phase.  }

\label{t:scalar}
$$\begin{array}{|c|c|c|c|c|c|c|}
\hline 

&W^{~} & \Psi^{~} , {\overline {\Psi}}^{~} & H^{~} & 
{\bar H}^{~}  & X^{~} &Z^{~} \\

\hline 
&&&&&&\\
{\cal R(Z)}  &3\alpha &\alpha & \frac{\alpha }{2} & 0 
&-\frac{\alpha }{2} &0   \\ 
& & & & & & \\
\hline
& & & & & & \\
{\cal U}(1)  &0 & 0 &0 &-\frac{17}{14}R  
&\frac{39}{14}R &R \\ 
& & & & & & \\
\hline
\end{array}$$
\end{table}

Including the lowest order `nonflat' terms
\footnote {The terms which contain 
$\Psi^3 $ or $\overline {\Psi}^3$ are `flat' and therefore do not affect 
the masses of the doublets and also do not take part in the fixing of the 
VEVs of the scalar fields. Because of this we do not take them into the 
account during the investigation of the higher order terms.}
the superpotential $W$ allowed  
by ${\cal R}({\cal Z})\times {\cal U}(1)$ symmetry is given by:

\begin{equation}
W=\Psi^3+\overline {\Psi }^3
+M_P^3\left(\frac{\overline {\Psi } \Psi }{M_P^2}\right)^5
+M_P^3\left(\frac{\bar HH}{M_P^2}\right)^{10}\left(\frac{X}{M_P}\right)^4
\frac{Z}{M_P}~.
\label{w}
\end{equation}
We easily observe
that along the directions (\ref {VEVs}) the SUSY conserving minima
of the potential is obtained for  $V=v=0$.
After  SUSY breaking 
{\em a' la} $N=1$ 
minimal supergravity theory, the
soft SUSY breaking mass terms which enter in the Lagrangian have the
form:

\begin{equation}  
V_{SSB}^m=m_{3/2}^2\left(|\Psi |^2+|\overline {\Psi }|^2+
|H|^2+|\bar H|^2 +|X|^2+|Z|^2\right)~,
\label{ssb}
\end{equation}  
where $m_{3/2}$ is the gravitino mass.
Together with (\ref {w}), one finds nonzero
solutions for $V$ and $v$ 
with the following magnitudes:

\begin{equation}  
V\sim M_P\left(\frac{m_{3/2}}{M_P} \right)^{1/8}~,~~~~
\langle X\rangle \sim \langle Z\rangle \sim
v\sim M_P\left(\frac{m_{3/2}}{M_P} \right)^{1/23}~.
\label{vevsol}
\end{equation}  
For $m_{3/2}= 10^3$~GeV and $M_P=2.4\cdot  10^{18}$~GeV 
(reduced Planck mass), 
we have:
$$
\epsilon \equiv \frac{V}{M_P}\sim 
\left(\frac{m_{3/2}}{M_P} \right)^{1/8}\sim 10^{-2}~,~~~~
$$
\begin{equation}
\epsilon_X\equiv \frac{v}{M_P}\sim \frac{\langle X\rangle }{M_P} 
\sim \frac{\langle Z\rangle}{M_P} \sim
\left(\frac{m_{3/2}}{M_P} \right)^{1/23}\sim 0.2~.
\label{vevsol1}
\end{equation}
Thus, at a scale of  $4.8\cdot 10^{17}$~GeV the 
$SU(6)\times U(1)$ group breaks to  $SU(5)\times U(1)'$ group through the
VEVs of the $H$, $\bar H$ fields. 
The VEVs ($V \sim 10^{16}$ GeV) of the $\Psi $, 
$\overline {\Psi }$ fields reduce the $SU(5)\times U(1)'$ symmetry 
to $SU(3)_C\times SU(2)_W\times U(1)_Y$. 

The lowest order $SU(6)^2\times U(1)$ global symmetry violating operator
$(\overline {\Psi }\Psi )^3(\bar HH)^9XZ^6$ 
which is permitted by the ${\cal R}\times {\cal U}(1)$ symmetry  
gives negligible
($\sim 1$~keV) contribution to the $\mu $-term.  
We have therefore obtained an `all order' solution of the gauge 
hierarchy problem and even an understanding of the origin of the GUT scale, 
whose magnitude is given by an interplay of
the two `fundamental' scales, the
Planck  
and the SUSY breaking scales.  Let us note that the properties of the 
scalar content of the $SU(3)^3$ gauge theory also permits one to understand 
the origin of the GUT scale \cite{gia1}.

While  the VEVs of the scalar
superfields obey the hierarchy $v\gg V$, taking into the account 
(\ref{gst}), (\ref{dst} )  
and (\ref{vevsol1}) we will see that the physical  doublet-antidoublets
reside in the $\overline {\Psi }+\Psi $ and $\bar H+H$ superfields 
respectively with the following weights:

$$
\overline {\Psi } \supset h_u~,~~~~~~~~~
H \supset \frac{\epsilon }{\epsilon_X }h_u~,
$$
\begin{equation}  
\Psi  \supset h_d~,~~~~~~~~~
\bar H \supset \frac{\epsilon }{\epsilon_X }h_d~.
\label{weights}
\end{equation}

\section{Charged Fermion Masses and Mixings }

In this section we will describe the pattern of charged fermion masses
and mixings in  our model. Together with the chiral supermultiplets
$\left(15+\bar 6_{-1}+\bar 6_1\right)^{(i)}$ ($i$ is a family index),
we introduce one pair of  $20_1+\overline {20}_{-1}$ which is necessary
for obtaining the top quark Yukawa coupling of order unity. 
Using this $20$-plet pair
the top quark mass is generated by the heavy particle exchange mechanism 
\cite{fro}.

Before considering all three generations let 
us demonstrate how the $b-\tau $
unification occurs in our scheme. 
The relevant couplings for down type
quark and lepton for the third family are:

\begin{equation}
W_Y(b, \tau )=\frac{1}{4\sqrt 2}A15^{(3)}15^{(3)} \Psi + 
\sqrt 2B\bar 6_{-1}^{(3)}\bar 6_1^{(3)}\Psi
+\sqrt 2C15^{(3)}\bar 6_{-1}^{(3)}\bar H~. 
\label{ybt}
\end{equation}

Here and below we use the proper normalization of the appropriate Yukawa
couplings.
Substituting the VEVs of the GUT and doublet fields the corresponding mass
matrices will be:

\begin{equation}
\begin{array}{cc}
 & {\begin{array}{cc} 
~~\bar g^{(3)} &\,\,~~d^{c(3)}
\end{array}}\\ \vspace{2mm}
\begin{array}{c}
q^{(3)} \\ g^{(3)}
\end{array}\!\!\!\!\! &{\left(\begin{array}{cc}
\,\, 0 &\,\,~~Ah_d
\\
 \,\, Cv &\,\,~~AV   

\end{array}\right)}~~,
\end{array}  \!\! ~~~~~
\begin{array}{cc}
 & {\begin{array}{cc} 
~~L^{(3)} &\,\,~~l^{(3)}  
\end{array}}\\ \vspace{2mm}
\begin{array}{c}
e^{c(3)} \\ \bar L^{(3)}    
 
\end{array}\!\!\!\!\! &{\left(\begin{array}{cc}
\,\, 0 &\,\,~~Bh_d   
\\ 
 \,\, Cv &\,\,~~ BV 

\end{array}\right) }~. 
\end{array}   ~~
\label{matbt}
\end{equation}
Assuming that $AV, BV \gg Cv$, the (22) elements of these matrices can 
be integrated out and we will have:
\begin{equation}
\lambda_b=\lambda_{\tau }=C\frac{v}{V}~.
\label{bt}
\end{equation}
Thus, in the framework of the flipped $SU(6)$ model we can 
obtain $b-\tau $
unification, which does not hold in  flipped $SU(5)$ models.

Assuming $C\sim 10^{-3}$
and for the values of $v$ and $V$ presented in 
(\ref{vevsol1}), we will have 
$\lambda_b=\lambda_{\tau }\sim 10^{-2}$, which suggests the regime  
$\tan \beta \sim 1$. This value of $\tan \beta $ is also 
preferred for the pseudo-Goldstone scenario \cite{tan} and proves useful 
for nucleon stability.  
The `smallness' of the C parameter is explained  below
by replacing it with the fourth power of the ratio $Z/M_P$.

Returning to three generations, together with the pair of 20-plets,
the Yukawa superpotential is given by:

$$
W_Y=\frac{1}{4\sqrt 2}{\cal A}_{ij}15^{(i)}15^{(j)}\Psi 
+\sqrt 2{\cal B}_{ij}\bar 6_{-1}^{(i)}\bar 6_{1}^{(j)}\Psi
+\sqrt 2{\cal C}_{ij}15^{(i)}\bar 6_{-1}^{(j)}\bar H 
+2{\cal D}_{ij}15^{(i)}\bar 6_{1}^{(j)}\frac{\overline {\Psi }H}{M_P}
$$
\begin{equation}
+\frac{1}{2\sqrt 3}{\cal F}_{i}15^{(i)}20_{1}H 
+\frac{1}{2\sqrt 3}{\cal G}_{i}\bar 6_{1}^{(i)}
\overline {20}_{-1}\overline {\Psi } 
+X\overline {20}_{-1}20_1~.
\label{wy}
\end{equation}
As mentioned above, the first three terms are responsible for generating
the masses of down quarks and leptons, while the remaining terms are 
relevant for the up type quarks. 
Let us note, that in our scheme we have to still impose the `ordinary'
family independent matter parity in order to guarantee nucleon stability.

For a reasonable explanation of the hierarchy of the 
Yukawa couplings we will assume
that the matrices ${\cal A}, \dots , {\cal D}, $ as well as the couplings 
${\cal F}$ and  ${\cal G}$ are not  constants but  operators which
depend on  powers of ratios $\bar HH/M_P^2$, $X/M_P$ and $Z/M_P$.
To avoid the unacceptable asymptotic relations 
$\lambda_s =\lambda_{\mu }$ and
$\lambda_d=\lambda_e$ in our scheme, the appropriate entries of 
the ${\cal B}$ matrix are  dependent on the ratio
$\bar HH/M_P^2$
\footnote{ In the framework of the `ordinary' $SU(6)$ pseudo-Goldstone
scenario, in the $3^{{\rm rd}}$ - $5^{{\rm th}}$ papers of ref. 
\cite{golds}  suggested ways of accommodating the fermion masses
and mixings.}.
The transformation 
properties  under ${\cal R}\times {\cal U}(1)$  
of the various matter superfields  
are presented in Table \ref{t:fer}.
The first four couplings of (\ref{wy}) can schematically  be 
written as:

\begin{table}
\caption{$R$ charges of the fermionic superfields
under the ${\cal R({\cal Z})}$ and ${\cal U}(1)$ symmetries.
}
\label{t:fer}
$$\begin{array}{|c|c|c|c|c|c|c|c|c|c|c|c|}
\hline 
 &15^{(1)}  & 15^{(2)}& 15^{(3)}&
\bar 6_{-1}^{(1)} &\bar 6_{-1}^{(2)} &
\bar 6_{-1}^{(3)}  & \bar 6_{1}^{(1)} &
\bar 6_{1}^{(2)}  &\bar 6_{1}^{(3)}
& 20_1 & \overline {20}_{-1} \\
\hline   
& & & & & & & & & & & \\
{\cal R(Z)}&\alpha  & \frac{3}{2}\alpha & \alpha &
\frac{5}{2}\alpha  &\frac{1}{2}\alpha  &\frac{3}{2}\alpha  &
 \frac{1}{2}\alpha &\frac{3}{2}\alpha  & \frac{1}{2}\alpha 
&2\alpha  &\frac{3}{2}\alpha \\
& & & & & & & & & & & \\
\hline 
& & & & & & & & & & & \\
{\cal U}(1) &-2R  &-\frac{39}{14}R  &0  &
-\frac{111}{14}R &\frac{17}{14}R  &-\frac{39}{14}R  &
-\frac{3}{14}R  &-\frac{39}{14}R &
\frac{39}{14}R&0 &-\frac{39}{14}R   \\
& & & & & & & & & & &\\
\hline
\end{array}$$
\end{table}

\begin{equation}
\begin{array}{ccc}
 & {\begin{array}{ccc} 
15^{(1)} &\,\,~~~~~15^{(2)} &\,\,~~15^{(3)}
\end{array}}\\ \vspace{2mm}
\begin{array}{c}
15^{(1)}\\ 15^{(2)} \\ 15^{(3)}
 
\end{array}\!\!\!\!\! &{\left(\begin{array}{ccc}
\,\, \left(\frac{Z}{M_P}\right)^4 &
\,\,~~ \frac{XZ^2}{M_P^3} &
\,\,~~ \frac{Z^2}{M_P^2}  
\\ 
\,\,  \frac{XZ^2}{M_P^3} &\,\,~~ 
\left(\frac{X}{M_P}\right)^2 &
\,\,~~ \frac{X}{M_P}    \\

\,\, \frac{Z^2}{M_P^2} &\,\,~~ \frac{X}{M_P} &\,\,~~1

 \end{array}\right)\cdot \Psi } 
\end{array}  \!\!~~~~~
\begin{array}{ccc}
 & {\begin{array}{ccc} 
~~~~~\bar 6_1^{(1)} &\,\,
~~~~~~~~~~\bar 6_1^{(2)}  &\,\,~~~\bar 6_1^{(3)}
\end{array}}\\ \vspace{2mm}
\begin{array}{c}
\bar 6_{-1}^{(1)} \\\bar 6_{-1}^{(2)}   \\\bar 6_{-1}^{(3)} 
 
\end{array}\!\!\!\!\! &{\left(\begin{array}{ccc}
\,\, \frac{Z\bar HH}{M_P^3}
\left(\frac{X}{M_P}\right)^3  &\,\,~~0 &\,\,~~0   
\\ 
\,\, 0 &\,\,~~ \frac{X\bar HH}{M_P^3} &\,\,~~0     \\

\,\, \left(\frac{Z}{M_P}\right)^3 &
\,\,~~ \left(\frac{X}{M_P}\right)^2 &\,\,~~1  

 \end{array}\right)\cdot \Psi } 
\end{array} ~~~
\label{matab}
\end{equation}

\begin{equation}
\begin{array}{ccc}
 & {\begin{array}{ccc} 
\hspace{-6mm}\bar 6_{-1}^{(1)}  &\,\,
~~~~~~~~~~\bar 6_{-1}^{(2)}  
&~~~~~~~~~~~\bar 6_{-1}^{(3)}
\end{array}}\\
\begin{array}{c}
15^{(1)}\\ 15^{(2)} \\ 15^{(3)}
 
\end{array}\!\!\!\!\! &{\left(\begin{array}{ccc}
\,\,\frac{X^3Z^4}{M_P^7}\frac{\bar HH}{M_P^2}  &\,\,~0 &
\,\,~\left( \frac{Z}{M_P}\right)^6
\\ 
\,\, \frac{X^4Z^2}{M_P^6}\frac{\bar HH}{M_P^2} &
\,\,~~\frac{\bar HH}{M_P^2}\left(\frac{Z}{M_P}\right)^4  &
\,\,~ \frac{X}{M_P}\left(\frac{Z}{M_P}\right)^4   \\

\,\,\frac{X^3Z^2}{M_P^5}\frac{\bar HH}{M_P^2}&\,\,~0 &
\,\,~~\left(\frac{Z}{M_P}\right)^4   

 \end{array}\right)\cdot \bar H } 
\end{array}  \!\!~  ~~~
\label{matc}
\end{equation}

\vspace{0.5cm}

\begin{equation}
\begin{array}{ccc}
 & {\begin{array}{ccc} 
\bar 6_1^{(1)}   &\,\,~~~~~~\bar 6_1^{(2)}  &\,\,~~~~\bar 6_1^{(3)}
\end{array}}\\ \vspace{6mm}
\begin{array}{c}
15^{(1)}\\ 15^{(2)} \\ 15^{(3)}
 
\end{array}\!\!\!\!\! &{\left(\begin{array}{ccc}
\,\, 0 &\,\,~~\frac{XZ^2}{M_P^3}  &\,\,~~~0   
\\ 
\,\, \left( \frac{Z}{M_P}\right)^3  &
\,\,~~\left(\frac{X}{M_P}\right)^2  &\,\,~~~1    \\

\,\,0&\,\,~~\frac{X}{M_P}  &\,\,~~~0     

 \end{array}~\right)\cdot \frac{\overline {\Psi }H }{M_P}} 
\end{array}  \!\!~.  ~~~
\label{matd}
\end{equation}
The ${\cal F}, {\cal G}$ terms have the form:
 
\begin{equation}
20_1\left(\frac{Z^2}{M_P^2}15^{(1)}+\frac{X}{M_P}15^{(2)}
+15^{(3)}\right)H +
\overline {20}_{-1}\left(\frac{Z^3}{M_P^3}\bar 6_1^{(1)}+
\frac{X^2}{M_P^2}\bar 6_1^{(2)}  
+\bar 6_1^{(3)} \right)\overline {\Psi }~.
\label{20}
\end{equation}
Without loss of generality one can choose the basis in which only
$15^{(3)}$ and $\bar 6_1^{(3)}$ states participate in 
(\ref{20}) .
Through the choice of this basis the couplings of 
(\ref{matab})-(\ref{matd}) will be the same.
So, for  estimates we  use couplings:
\begin{equation}
\frac{F}{2\sqrt 3}20_115^{(3)}H +
\frac{G}{2\sqrt 3}
\overline {20}_{-1}\bar 6_1^{(3)} \overline {\Psi }~.
\label{20up}
\end{equation}
These couplings are relevant for the generation of the top quark mass
and should be taken into account during the analyses of the neutrino 
masses as well.
For 
$\epsilon $ and $\epsilon_X $ (see (\ref{vevsol1}))  
we will take the values $10^{-2}$ and 
$0.2$ respectively.

Starting with the masses of the charged leptons the mass matrix
relevant for the three families will have the form:

\begin{equation}
\begin{array}{cccccc}
 & {\begin{array}{cccccc} 
~~~~L^{(1)} &
 \,\,~~~~~~ L^{(2)} &
 \,\,~~~~~~L^{(3)} &
\,\, ~~~~~~l^{(1)}&\,\,~~~~~~~~l^{(2)}&
\,\,~~~~~~l^{(3)}
\end{array}}\\ \vspace{2mm}
\hat{M}_L= \begin{array}{c}
e^{c(1)} \\e^{c(2)}  \\e^{c(3)} 
\\ \bar L^{(1)}  \\\bar L^{(2)}  \\
 \bar L^{(3)}
\end{array}\!\!\!\!\! &{\left(\begin{array}{cccccc}
\,\, 0 &\,\,~~0 &\,\,~~0  &\,\,~~b_{11}'\epsilon_X^6h_d &\,\,~~0&\,\,~~0 
\\ 
\,\, 0 &\,\,~~0 &\,\,~~0 &\,\,~~0 &\,\,~~b_{22}'\epsilon_X^3 h_d&\,\,~~0  
\\
\,\,0&\,\,~~0 &\,\,~~0 &\,\,~~b_{31}\epsilon_X^3h_d &
\,\,~~b_{32}\epsilon_X^2h_d&\,\,~~b_{33}h_d
\\
\,\,c_{11}v\epsilon_X^9 &\,\,~~c_{21}v\epsilon_X^8 &
\,\,~~c_{31} v\epsilon_X^7 &\,\,~~b_{11}V\epsilon_X^6&\,\,~~0 &\,\,~~0 
\\ 
\,\,0&\,\,~~c_{22}v\epsilon_X^6  &\,\,~~0&\,\,
~~0& \,\,~~b_{22}V\epsilon_X^3 &\,\,~~0
\\
\,\,c_{13}v\epsilon_X^6 &\,\,~~c_{23}v\epsilon_X^5&
\,\,~~c_{33}v\epsilon_X^4  &
\,\,~~b_{31}\epsilon_X^3 
&\,\,~~b_{32}V\epsilon_X^2 &\,\,~~b_{33}V
 \end{array}\right)} 
\end{array}  \!\!  ~~~~~
\label{bigmate}
\end{equation} 
where coefficients with primes appear because 
there are several possibilities
of the convolution of the gauge indices in the 
corresponding entries of the
second `matrix term' of (\ref{matab}).
One can easily see that for  values of the constants $c_{21}\sim 1/3$ 
and $c_{31}\sim 1/10$ the see-saw limit works well and the states 
which correspond to the $3\times 3$ block of the lowest right 
side  of the (\ref{bigmate}) matrix could be integrated out. 
For the `light' charged leptons
we obtain the following mass matrix:

\begin{equation}
\begin{array}{ccc}
 & {\begin{array}{ccc} 
l_1~~~~~~ & \,\,~~~~l_2~~~~~ & \,\,~~~~l_3~~~~~~~~
\end{array}}\\ \vspace{2mm}
\hat{m}_e= \begin{array}{c}
e^c_1 \\e^c_2  \\e^c_3 
 \end{array}\!\!\!\!\! &{\left(\begin{array}{ccc}
\,\,c_{11}\frac{b_{11}'}{b_{11}}\epsilon_X^5  &
\,\,~~c_{21}\frac{b_{11}'}{b_{11}} \epsilon_X^4 &
\,\,~~c_{31}\frac{b_{11}'}{b_{11}} \epsilon_X^3 
\\ 
\,\,0  &\,\,~~c_{22}\frac{b_{22}'}{b_{22}} \epsilon_X^2  &\,\,~~0

\\
\,\,c_{13}\epsilon_X^2 &
\,\,~~c_{23}\epsilon_X  &\,\,~~c_{33} 
\end{array}\right) \frac{\epsilon_X^5}{\epsilon } h_d}~. 
\end{array}  \!\!  ~~~~~
\label{mate}
\end{equation} 
A biunitary transformation

\begin{equation}
R_e^{\dag }\hat{m}_eL _e=\hat{m}_e^{diag}~,~~~
\nonumber \\
\begin{array}{ccc}
L_e=~~ \\
\end{array}
\hspace{-6mm}\left(
\begin{array}{ccc}
1& 0 & s_{13}^l \\
0& 1& s_{23}^l \\
-s_{13}^l&-s_{23}^l & 1 \end{array}
\right),~~~~
\label {Le}
\end{equation}
where
\begin{equation}
s_{13}^l=\frac{c_{13}}{c_{33}}\epsilon_X^2~,~~~
s_{23}^l=\frac{c_{23}}{c_{33}}\epsilon_X~,~~~
\label{sl}
\end{equation}
transforms the matrix in (\ref{mate}) into the diagonal form with
eigenvalues:
\begin{equation}
\lambda_e = 
c_{11}\frac{b_{11}'}{b_{11}} \frac{\epsilon_X^{10} }{\epsilon }~,~~~~
\lambda_{\mu }=
 c_{22}\frac{b_{22}'}{b_{22}}\frac{\epsilon_X^7}{\epsilon }~,~~~~
\lambda_{\tau }=c_{33}\frac{\epsilon_X^5}{\epsilon }~,
\label{lep}
\end{equation}
In estimating quark mixings and nucleon decay we will 
deal with $L$ matrices which rotate the basis of the left handed fields.
In obtaining (\ref{lep}) 
the smallness of $c_{21}$ and $c_{31}$ with respect to the
other couplings has been taken into account.

Following the same strategy the mass matrix which 
leads to the masses of down quarks 
will have the form:

\vspace{0.5cm}
\begin{equation}
\begin{array}{cccccc}
 & {\begin{array}{cccccc} 
\,\,~~~\bar g^{(1)} &
 \,\,~~~~~~ \bar g^{(2)} &
 \,\,~~~~~~\bar g^{(3)}&
\,\,~~~~~~~d^{c(1)} &\,\,~~~~~~~ d^{c(2)}&
\,\,~~~~~~d^{c(3)}
\end{array}}\\ \vspace{2mm}
\hat{M}_D= \begin{array}{c}
q^{(1)}\\ q^{(2)} \\ q^{(3)}
\\g^{(1)} \\g^{(2)}\\g^{(3)} 
\end{array}\!\!\!\!\! &{\left(\begin{array}{cccccc}
\,\, 0 &\,\,~~0 &\,\,~~0  &\,\,~~a_{11} \epsilon_X^4h_d 
&\,\,~~a_{12}\epsilon_X^3h_d&\,\,~~a_{13}\epsilon_X^2h_d 
\\ 
\,\, 0 &\,\,~~0 &\,\,~~0 &\,\,~~a_{12}\epsilon_X^3 h_d 
&\,\,~~ a_{22}\epsilon_X^2 h_d&\,\,~~a_{23}\epsilon_Xh_d  
\\
\,\,0&\,\,~~0 &\,\,~~0 &\,\,~~a_{13}\epsilon_X^2h_d  &
\,\,~~a_{23}\epsilon_X h_d&\,\,~~a_{33}h_d
\\
\,\,c_{11}v\epsilon_X^9 &\,\,~~0 &\,\,~~c_{13}v\epsilon_X^6 &
\,\,~~a_{11}V\epsilon_X^4 
&\,\,~~a_{12}V\epsilon_X^3 &\,\,~~a_{13}V\epsilon_X^2
 \\ 
\,\,c_{21}v\epsilon_X^8 &\,\,~~c_{22}v\epsilon_X^6  &
\,\,~~c_{23}v\epsilon_X^5 &\,\,
~~a_{12}V\epsilon_X^3 & \,\,~~a_{22}V\epsilon_X^2 &\,\,~~a_{23}V\epsilon_X 
\\
\,\,c_{31}v\epsilon_X^7  &\,\,~~ 0 &\,\,~~c_{33}v\epsilon_X^4 &
\,\,~~a_{13}V\epsilon_X^2 
&\,\,~~a_{23}V\epsilon_X  &\,\,~~a_{33}V
 \end{array}\right)} 
\end{array}  \!\!  ~~~~~
\label{bigmatd}
\end{equation} 
and, after integrating out the appropriate heavy states,
this matrix reduces to: 
\begin{equation}
\begin{array}{ccc}
 & {\begin{array}{ccc} 
\hspace{-5mm}~d^c_1~~ & \,\,~~~~d^c_2~~  & \,\,~~~~d^c_3~~~
\end{array}}\\ \vspace{2mm}
\hat{m}_d= \begin{array}{c}
q_1\\ q_2 \\q_3 
 \end{array}\!\!\!\!\! &{\left(\begin{array}{ccc}
\,\,c_{11}\epsilon_X^5  &\,\,~~0 &\,\,~~c_{13}\epsilon_X^2 
\\ 
\,\,c_{21}\epsilon_X^4   &\,\,~~c_{22}\epsilon_X^2  &\,\,~~c_{23}\epsilon_X
 \\
\,\, c_{31}\epsilon_X^3 &\,\,~~ 0  &\,\,~~c_{33} 
\end{array}\right) \frac{\epsilon_X^5}{\epsilon } h_d}~. 
\end{array}  \!\!  ~~~~~
\label{matdown}
\end{equation} 
By a transformation:
\begin{equation}
\nonumber \\
L_d^{\dag }\hat{m}_dR_d=\hat{m}_d^{diag},~~~~~
\begin{array}{ccc}
L_d=~~ \\
\end{array}
\hspace{-6mm}\left(
\begin{array}{ccc}
1&  0 & s_{13}^d \\
0& 1& s_{23}^d \\
-s_{13}^d&-s_{23}^d & 1 \end{array}
\right),~~~~
\label {Ld}
\end{equation}
where
\begin{equation}
s_{13}^d= \frac{c_{13}}{c_{33}}\epsilon_X^2 ~,
~~~~~~s_{23}^d= \frac{c_{23}}{c_{33}} \epsilon_X~, 
\label{sd}
\end{equation}
the matrix in (\ref {matdown}) is diagonalized with eigenvalues 
given by:
\begin{equation}
\lambda_d= c_{11}\frac{\epsilon_X^{10} }{\epsilon }~,~~~~
\lambda_s= c_{22}\frac{\epsilon_X^7}{\epsilon }~,~~~~
\lambda_b= c_{33}\frac{\epsilon_X^5}{\epsilon }~.
\label{down}
\end{equation}

Turning to the masses of the up type quarks from the couplings 
(\ref{matd}), (\ref{20up}) 
and also 
taking into account the last term of (\ref{wy}), 
the
matrix which is responsible for generation
of the masses of the up quarks will have the form:
\begin{equation}
\begin{array}{cccc}
 & {\begin{array}{cccc} 
\hspace{-0.3cm}~~~~~~~~u^{c(1)} &
 \,\,~\hspace{0.8cm}u^{c(2)}  &
 \,\,~~~\hspace{0.5cm}u^{c(3)} &
\,\,~\hspace{0.5cm}\bar q_{20}
\end{array}}\\ \vspace{2mm}
\hat{M}_U= \begin{array}{c}
q^{(1)} \\q^{(2)}  \\ q^{(3)}
\\ q_{\overline {20}}
\end{array}\!\!\!\!\! &{\left(\begin{array}{cccc}
\,\, 0 &\,\,~~d_{12}\epsilon_X^4h_u &\,\,~~0  &
\,\,~~0 
\\ 
\,\,d_{21}\epsilon_X^4h_u  &\,\,~~d_{22}\epsilon_X^3h_u &
\,\,~~d_{23}\epsilon_Xh_u &
\,\,~~0
\\
\,\,0&\,\,~~d_{32}\epsilon_X^2h_u  &\,\,~~0 &
\,\,~~Fv
\\
\,\,0 &\,\,~~0 &\,\,~~Gh_u &\,\,~~\langle X \rangle

\end{array}\right)} 
\end{array}  \!\!~  ~~~~~
\label{bigmatu}
\end{equation} 
Integrating out the corresponding heavy fragments from the 20-plets 
we obtain 
the following up quark mass matrix:

\begin{equation}
\begin{array}{ccc}
 & {\begin{array}{ccc} 
\,\,\hspace{-0.4cm}u^c_1 &
 \,\,~~\hspace{0.5cm}u^c_2  &
 \,\,~~~\hspace{0.5cm}u^c_3 
\end{array}}\\ \vspace{2mm}
\hat{m}_u= \begin{array}{c}
q_1 \\q_2  \\ q_3
\end{array}\!\!\!\!\! &{\left(\begin{array}{ccc}
\,\, 0 &\,\,~~d_{12}\epsilon_X^4&\,\,~~0  
\\ 
\,\,d_{21}\epsilon_X^4 &\,\,~~d_{22}\epsilon_X^3 &\,\,~~d_{23}\epsilon_X 
\\
\,\,0&\,\,~~d_{32}\epsilon_X^2 &\,\,~~\lambda_t 
\end{array}\right)h_u} 
\end{array}  \!\!  ~~~~~
\label{matu}
\end{equation} 
Rotating the basis of the left-right fields
\begin{equation}
L_u^{\dag }\hat{m}_uR_u=\hat{m}_u^{diag},~~~~~
\nonumber \\
\begin{array}{ccc}
L_u=~~ \\
\end{array}
\hspace{-6mm}\left(
\begin{array}{ccc}
1& s_{12}^u & 0 \\
-s_{12}^u& 1& s_{23}^u \\
0&-s_{23}^u & 1 \end{array}
\right),~~~~
\nonumber \\
\label {Lu}
\end{equation}
where
\begin{equation}
s_{12}^u=\frac{d_{12}}{\lambda_c}\epsilon_X^4\sim \epsilon_X~,
~~~s_{23}^u=\frac{d_{23}}{\lambda_t} \epsilon_X~.
\label{su}
\end{equation}
we will have:
\begin{equation}
\lambda_t =-FG\frac{v}{\langle X \rangle } \sim 1~,~~~
\lambda_c \sim \epsilon_X^3~,
~~~\lambda_u \sim \epsilon_X^5~,
\label{up}
\end{equation}

We see that by   imposing the  ${\cal R}\times {\cal U}(1)$ 
horizontal symmetry
we can obtain a reasonable 
description of fermion mass hierarchy. 
The $b-\tau $ unification still holds, while the unwanted
relations $\lambda_s=\lambda_{\mu }$ and $\lambda_d=\lambda_e$ 
are avoided. 
The masses of the electron and 
$\mu $ meson are proportional respectively to 
$\frac{b_{11}'}{b_{11}}$ and $\frac{b_{22}'}{b_{22}}$, while the down quark
mass matrix depends only on the $c_{ij}$ couplings (see (\ref{matdown})).
Since appropriate couplings (see (\ref{matab})) are built through the
nonrenormalizable operators and there exist several ways of the 
convolution of the gauge indices, the $b$, $b'$ constants are completely
arbitrary and 
taking $\frac{b_{11}'}{b_{11}}\sim \frac{b_{22}}{b_{22}'}\sim 1/3$
we will have

\beq
m_s\sim \frac{1}{3}m_{\mu }~,~~~~~~~~
\frac{m_s}{m_d}\sim \frac{1}{9}\frac{m_{\mu }}{m_e}\sim 20~,
\label{asympt}
\eeq
which are desirable asymptotic relations.

Although the masses of quark-lepton families do not depend  on the
$a_{ij}$ coefficients, limits on their magnitudes could be 
obtained from some physical considerations.  
These coefficients could not be taken very small because
the masses of the decoupled (heavy) states crucially depend on their 
magnitudes.
Taking them too small the see-saw limit in which the 
matrices (\ref{mate}) and (\ref{matdown})  
were
obtained does not work, and formulae
(\ref{lep}) and (\ref{down}) are invalid.
 The masses of the (decoupled) three generations of triplet and doublet 
states are respectively:

$$
m_T^{(3)}=a_{33}V~,~~~~~~~~~~~m_D^{(3)}=b_{33}V~,
$$
$$
m_T^{(2)}=(a_{22}-\frac{a_{23}^2}{a_{33}})\epsilon_X^2V~,~~~~~~~~~
m_D^{(2)}=b_{22}\epsilon_X^3V~,
$$
\begin{equation}
m_T^{(1)}=(a_{11}-\frac{a_{13}^2}{a_{33}}-
\frac{a_{12}^2}{a_{22}-a_{23}^2/a_{33}})\epsilon_X^4V~,~~~
m_D^{(1)}=b_{11}\epsilon_X^6V~.
\label {massdt}
\end{equation}
From these formulae and  for values of the parameters:
$$
a_{33}\sim b_{11}\sim b_{22}\sim b_{33}\sim 1~, 
$$
$$
a_{11}\sim \epsilon_X^2~,~~~a_{22}\sim \epsilon_X
$$
\begin{equation}
a_{12}\sim (\epsilon_X)^{3/2}~,~~~
a_{13}\sim \epsilon_X~,~~~
a_{23}\sim (\epsilon_X)^{1/2}~,
\label{as}
\end{equation}
we will have
\begin{equation}
m_T^{(i)}\simeq m_D^{(i)}~,
\label{dt}
\end{equation}
and the successful unification of the three gauge couplings 
will be unchanged  in the one loop level.
For  values of the coefficients in  (\ref{as}) the see-saw limit
still works if $c_{13}\sim 2\cdot 10^{-2}$ and $c_{23}\sim 0.2$. 
Note that the picture will not be changed if 
the coefficients $a_{12}$, $a_{13}$
and $a_{23}$ are taken to be less than the values in (\ref{as}). 
As we will see in
the next section, the proton decay rate in our model crucially 
depends on these
parameters and we can say more about them from the requirement 
of proton stability as demanded by experiments.

\vspace{0.5cm}

To conclude this section, let us summarize the results which we have  obtained.
For the Yukawa couplings and hierarchies in the down quark and lepton
sectors we have:
$$
\lambda_b =\lambda_{\tau }\sim \frac{\epsilon_X^5}{\epsilon}~,~~~
\tan \beta \sim 1~,
$$
$$
\lambda_d :\lambda_s :\lambda_b \sim \epsilon_X^5:\epsilon_X^2 :1~,
$$
\begin{equation}
\lambda_e :\lambda_{\mu } :\lambda_{\tau } \sim \epsilon_X^5:\epsilon_X^2 :1~,
\label{lamdaed}
\end{equation}
while for the up quarks:
$$
\lambda_t \sim  1~,
$$
\begin{equation}
\lambda_u : \lambda_c :  \lambda_t \sim 
\epsilon_X^5 : \epsilon_X^3 :1~, 
\label{lamdau}
\end{equation}
In the notation used above (see (\ref{Ld}) and (\ref{Lu})) 
the CKM matrix is:
\begin{equation}
\hat{V}=L_u^{T}L_d^{*}  
\label {ckm}
\end{equation}
with the matrix elements
$$
\hat{V}_{us}=\frac{d_{12}}{\lambda_c}\epsilon_X^4\sim \epsilon_X~,
$$
$$
\hat{V}_{ub}=\left(c_{12}^{*}\epsilon_X-\hat{V}_{us}c_{23}^{*}\right)
\frac{\epsilon_X}{c_{33}^{*}}~,
$$
\begin{equation}
\hat{V}_{cb}=\left(\frac{c_{23}^{*}}{c_{33}^{*}}-
\frac{d_{23}}{\lambda_t}\right)\epsilon_X~.
\label{ckmels}
\end{equation}
With
$c_{23}\sim d_{23}\sim 1/5$, $c_{13}\sim 2\cdot 10^{-2}$ 
and the other participating couplings 
of order unity, the CKM matrix elements all have acceptable values
\footnote{As we will see in the next section for proton stability a
better value for $c_{23}$ is $\sim 1/12$. In this case
$V_{ub}$ and  $V_{cb}$ can still  have the desired values since they
also contain other entries.}.

\section{Nucleon Decay}

Dimension five nucleon decay occurs in our model through exchange  
of the colored Higgsinos from the $\Psi $ 
and $\overline {\Psi }$ superfields. The colored states from the $\bar H+H$
superfields are goldstones `eaten' by the appropriate gauge fields and are 
irrelevant for nucleon decay. In our model there exist insertions 
(see (\ref{wm})) between triplet states 
which come from the same superfields (from $\Psi$
and $\overline {\Psi }$). Because of this the $d=5$ operator which 
is related to the left handed fields can emerge only from the first
coupling in (\ref{wy}). Its decomposition into components which are relevant
for nucleon decay is:
\begin{equation}
\frac{1}{4\sqrt 2}{\cal A}_{ij}15_{(i)}15_{(j)}\Psi \to
\frac{1}{2}{\cal A}_{ij}q_iq_j g_{\Psi}+
{\cal A}_{ij}q_iL_jd^c_{\Psi}~.
\label{d5}
\end{equation}

From the mass matrix in (\ref{bigmate}) we have seen that 
left handed states of the 
lepton doublets mainly come from the $L_{i}$ states.
Taking into account (\ref{Le}), (\ref{Ld}) and (\ref{Lu})  
in the mass eigenstate basis the couplings  (\ref{d5})
take the form:

\begin{equation}
u(L_u^{\dag  }{\cal A}L_d^{*})dg_{\Psi }+
u(L_u^{\dag }{\cal A}L_e)ed^c_{\Psi } -
d(L_d^{\dag }{\cal A}L_e)\nu d^c_{\Psi } ~,
\label {d51}
\end{equation}
After integrating out the heavy triplet states and dressing the $d=5$
operators by the wino states, we will obtain
the four fermion operators for 
nucleon decay. The neutrino decay channel will occur through the operator:
\begin{equation}
{\cal O}=x\cdot (u^a d^b_m)(d^c_j\nu_i)\varepsilon_{abc}
\label{op}
\end{equation}
where  $a, b, c$ are color indices and
$$
x=\alpha \left(
-(L_d^{\dag }{\cal A}L_e)_{ji}(L_u^{\dag }{\cal A} L_d^{*})_{ln}
\hat{V}_{lm}\hat{V}^{\dag }_{n1}
+
(L_u^{\dag }{\cal A} L_d^{*})_{1m}(L_u^{\dag }{\cal A} L_e)_{ln}\hat{V}_{lj}-
\right.
$$
\begin{equation}
\left.
~~~~~~~-(L_d^{\dag }{\cal A} L_u^{*})_{jl}
(L_d^{\dag }{\cal A}L_e)_{ni}\hat{V}_{lm} \hat{V}^{\dag }_{n1}
-
(L_d^{\dag }{\cal A} L_u^{*})_{ml}
(L_u^{\dag }{\cal A} L_e)_{1i}\hat{V}_{lj}
\right)
\label {x0}
\end{equation}
where $\alpha $ is a family independent factor.

For nucleon decay the first two generations of  
quarks and leptons occur in the external lines,
while the contribution of the third
generation through internal loops are somewhat suppressed.
For estimates 
we will assume that the family indices can be $1$ and/or $2$. 
Since for two generations the only mixing term of the 
CKM matrix is the Cabbibo
angle, which is not renormilized between the GUT and electroweak scales 
\cite{ren} ,
we can replace $\hat{V}$ in (\ref{x0}) using formula (\ref{ckm}). 
After substituting $\hat{V}$ and summing over
the repeated indices we see that the first two terms of  (\ref{x0})
exactly cancel out, while the
sum of  the remaining two terms gives:   
\begin{equation}
x=-2\alpha 
(L_d^{\dag }{\cal A} L_d^{*})_{mj}(L_u^{\dag }{\cal A} L_e)_{1i}~.
\label{x01}
\end{equation}

Similarly the four fermion operator which corresponds to nucleon
decay into charged leptons is given by:

\begin{equation}
{\cal O'}=x'\cdot (u^a d^b_j)(u^ce_i)
\varepsilon_{abc}    
\label{op1}
\end{equation}
with
$$
x'=\alpha \left(-(L_u^{\dag }{\cal A} L_d^{*})_{1j}
(L_d^{\dag }{\cal A} L_e)_{mi}\hat{V}^{\dag }_{m1}
+
(L_u^{\dag }{\cal A} L_e)_{1i}
(L_u^{\dag }{\cal A} L_d^{*})_{mn}\hat{V}_{mj}\hat{V}^{\dag }_{n1}+
\right.
$$
\begin{equation}
\left.
~~~~~~~~~~~+(L_u^{\dag }{\cal A} L_d^{*})_{1m}
(L_d^{\dag }{\cal A} L_e)_{ji}\hat{V}^{\dag }_{m1}
+
(L_u^{\dag }{\cal A} L_d^{*})_{1m}
(L_u^{\dag }{\cal A} L_e)_{ni}\hat{V}_{nj}\hat{V}^{\dag }_{m1}
\right)
\label{x1}
\end{equation}
After some simplifications we find:
\begin{equation}
x'=2\alpha (L_u^{\dag }{\cal A} L_u^{*})_{11}(L_d^{\dag }{\cal A} L_e)_{ji}
\label{x11}
\end{equation}

From (\ref{op}) and (\ref{x01}), the $p\to K\nu_{\mu }$ decay 
width normalized with respect to the $SU(5)$ case  
\cite{mur} is:
\begin{equation}
\frac{\Gamma (p\to K\nu_{\mu })}{\Gamma_{SU(5)}(p\to K\nu_{\mu })}=
\left[\frac{a_{12}(a_{22}\sin \theta +a_{12}\epsilon_X)\epsilon_X^5}
{\lambda_s \lambda_c \sin ^2\theta }\right]^2~.
\label{pnu}
\end{equation}
Similarly, the decay width for the reaction $p\to K\mu $ is:
\begin{equation}
\frac{\Gamma (p\to K\mu )}{\Gamma_{SU(5)}(p\to K\nu_{\mu })}=
0.12\cdot \left[\frac{a_{22}(a_{11}\epsilon_X^2-
2a_{12}\epsilon_X\sin \theta +a_{22}\sin^2 \theta )\epsilon_X^4}
{\lambda_s \lambda_c \sin ^2\theta }\right]^2~,
\label{pmu}
\end{equation}
where the factor $0.12$ arises from the difference 
between proton-neutrino and proton-charged lepton hadronic
matrix elements
\cite{mur}.

\begin{table}
\caption{Proton  lifetime in  units
of $\tau (p\to K\nu_{\mu })_{SU(5)}$.}
\label{t:tab1}
$$\begin{array}{|c|c|c|c|}
\hline
 & ({\it i}) & ({\it ii}) & ({\it iii})
\\
\hline
\tau (p\to K\nu_{\mu }) &\sim  10^2
&\stackrel{_>}{_\sim} 10^3
&\sim 6\cdot 10^2
\\
\hline
\tau (p\to K\nu_e) &\sim 2\cdot 10^4
&\stackrel{_>}{_\sim} 8\cdot10^5
& \sim 2\cdot 10^4
\\
\hline
\tau (p\to \pi \nu_{\mu }) &\sim 7\cdot 10^3
&\sim 7\cdot10^3  & \sim 3\cdot 10^4
\\
\hline
\tau (p\to \pi \nu_e) &\sim 8\cdot 10^5
&\sim 4\cdot10^6  & \sim 8\cdot 10^5
\\
\hline
\hline
\tau (p\to K\mu ) &\sim 2\cdot 10^2
&\sim 2\cdot 10^2  & \sim 5\cdot 10^3
\\
\hline
\tau (p\to Ke ) &\sim 3\cdot 10^4
&\sim 10^5  & \sim  10^5
\\
\hline
\tau (p\to \pi \mu ) &\sim 3\cdot 10^3
&\sim 10^4  & \sim  10^4
\\
\hline
\tau (p\to \pi e ) &\sim 3\cdot 10^5
&\sim 3\cdot 10^5  & \sim 2\cdot 10^6
\\
\hline
\end{array}$$
\end{table}

As we mentioned earlier, there is some freedom
in the choice of the $a_{ij}$ parameters. None of them should 
exceed the values presented in (\ref{as}) in order to keep unification 
of three gauge coupling constant at $M_{GUT}$. 
For appropriate $a_{ij}$ given by (\ref{as}) the decay widths in
(\ref{pnu}) and (\ref{pmu}) will have the values:

\begin{equation}
\frac{\Gamma (p\to K\nu_{\mu })}{\Gamma_{SU(5)}(p\to K\nu_{\mu })}
\simeq 7.7\cdot 10^{-3}~,~~~~~~
\frac{\Gamma (p\to K\mu )}{\Gamma_{SU(5)}(p\to K\nu_{\mu })}
\simeq  5\cdot 10^{-3}~,
\label{pnu1}
\end{equation}
which are well suppressed relative  to the  dominant $SU(5)$
decay mode. 
However, as we will see below, there are parameter choices for which 
the  charged lepton decay mode is dominant.

Since  
(\ref{pnu}), (\ref{pmu}) are sensitive to the couplings $a_{12}$
and $a_{22}$ we will study proton decay by 
varying $a_{12}$ and $a_{22}$, keeping the  other $a_{ij}$ 
unchanged. Three cases, which give different phenomenological
implications for proton decay, will be relevant:

$$
({\it i})~a_{22}\sim \epsilon_X~,~~~a_{12}\sim \epsilon_X^{3/2}~;~~~
({\it ii})~a_{22}\sim \epsilon_X~,~~~a_{12}\stackrel {_<}{_\sim }
0.3\cdot \epsilon_X^{3/2}~;
$$
\beq
({\it iii})~a_{22} \stackrel {_<}{_\sim }  \epsilon_X^{3/2}~,~~~
a_{12} \sim \epsilon_X^{3/2}~.
\label{spaces}
\eeq

As we have seen, $({\it i})$ gives the same proton decay 
rate in the $K\nu_{\mu }$  and $K\mu $ channels. 
However, in  case $({\it ii})$ of
(\ref{spaces}) the reaction $p\to K\mu $ dominates
\footnote{Refs. 
\cite{ptokmu} also discuss models which also
predict nucleon decay with emission of the charged lepton.}, 
while case $({\it iii})$
has dominant decay $p\to K\nu_{\mu }$. In Table (\ref{t:tab1}) we
present the proton lifetimes in the units of
$\tau (p\to K\nu_{\mu })_{SU(5)}$. As we see the other decay
channels are more suppressed for all three cases.


In conclusion, we expect the proton to be quite long lived 
($\tau_p \stackrel {_>}{_\sim }10^{33\pm 2}$~yr.).

\subsection{Suppression of Planck Scale Induced $d=5$ Operators}

Even in the framework of the minimal SUSY standard model Planck-scale
physics may generate $d=5$ operators
\begin{equation}
\frac{\lambda }{M_P}
q\cdot q\cdot q\cdot l~,
\label{d5pl}
\end{equation}
which are permitted by the matter $R$- parity,
and could induce rapid  nucleon
decay if the dimensionless coupling $\lambda $ 
is not less then
$10^{-8}$ or so. It is therefore desirable to 
have some mechanism which naturally
suppress such operators.
One possibility to forbid them is to employ discrete gauge symmetries
\cite{iba}. Other ways include a  
prescription, where the matter fields have family dependent transformation
properties under some flavor group \cite{d5hor} ,
or to have a `redefined' $R$-symmetry. The last 
possibility was suggested recently \cite{d5r} where all baryon number 
violating terms were forbidden.

Since in our $SU(6)\times U(1)$ model the matter superfields carry 
family dependant ${\cal R}\times {\cal U}(1)$ charges the 
suppression of the Planck scale $d=5$ operators 
can be checked directly.

The relevant terms include:
 
\begin{equation}
\frac{1}{M_P^2}
\hat{\Gamma }_{ijmn}
\cdot 15^{(i)}\cdot 15^{(j)}\cdot 15^{(m)}\cdot 15^{(n)}
\cdot \overline {\Psi }~,
\label{d50}
\end{equation}

\begin{equation}
\frac{1}{M_P^2}\hat{\Gamma }^{'}_{ijmn}
\cdot 15^{(i)}\cdot 15^{(j)}\cdot 15^{(m)}
\cdot \bar 6_1^{(n)}\cdot H~,
\label{d52}
\end{equation}
where $i,\dots ,n$ are family indices and
$\hat{\Gamma }$, $\hat{\Gamma }^{'}$ are tensors which depend 
on powers of $\overline {\Psi }\Psi $, $\bar HH$, $X$ and $Z$ 
superfields, and are chosen in such a way as to respect the 
${\cal R}\times  {\cal U}(1)$ symmetry.

The elements of $\hat{\Gamma }$ 
which correspond to the processes $p\to K\nu_{e, \mu , \tau }$  
respectively are:

$$
\hat{\Gamma }_{1112}=\frac{\bar HH}{M_P^2}
\left(\frac{\overline \Psi \Psi }{M_P^2} \right)^2
\left(\frac{Z}{M_P} \right)^{10}~,~~~~~~
\hat{\Gamma }_{1122}=\frac{\bar HH}{M_P^2}\frac{X}{M_P}
\left(\frac{\overline \Psi \Psi }{M_P^2} \right)^2
\left(\frac{Z}{M_P} \right)^8~,
$$
\begin{equation}
\hat{\Gamma }_{1123}=\frac{\bar HH}{M_P^2}
\left(\frac{\overline \Psi \Psi }{M_P^2} \right)^2
\left(\frac{Z}{M_P} \right)^8~.
\label{gamma}
\end{equation}
Substituting the VEVs of the scalar superfields and taking 
(\ref {d50}) into the account,
we see that the  appropriate $d=5$ operators :

\begin{equation}
\frac{1}{M_P}~\epsilon^5~\epsilon_X^{10}~q_1q_1q_2
(\epsilon_X^2l_1+\epsilon_Xl_2+l_3)
\label{d5nu}
\end{equation}
are strongly suppressed. Building all the other $\hat{\Gamma }$ `coefficients'
we can check that they all contain at least a factor 
($\sim \epsilon^4$),
which is already enough for the required suppression. 

In (\ref {d52}), the $SU(2)_W$ doublet state should be extracted
from $\bar 6_1^{(n)}$ . The latter  are the decoupled heavy states
and contain the light $l_i$ doublets given by the weights 
$\epsilon_X^{~k_i}/\epsilon $ , where $k_1=k_2=4$, $k_3=5$.
As far as the $\hat{\Gamma}^{'}$s are concerned, they contain at least  
the $(\overline \Psi \Psi )^2$ combination, which makes nucleon decay
unobservable. 

Finally, note, that in building the fermion mass sector we 
introduced a pair
of $20$-plets for the generation of the top quark mass. 
Since the $\overline {20}$ state contains a $q_3$ state of weight 1, 
one should
look for terms which could induce the relevant $d=5$ operators. 
Writing down all possible operators:

\begin{equation}
\frac{1}{M_P}\hat{\Omega }_{ijm}
\cdot \overline {20}\cdot 15^{(i)}\cdot 15^{(j)}
\cdot \left(\bar 6_1^{(m)}+\hat{\omega } ~
\frac{\overline \Psi \bar H}{M_P^2}~ 15^{(m)}\right)~,
\label{d53}
\end{equation}

\begin{equation}
\frac{1}{M_P^2}\hat{\Upsilon }_{ij}
\cdot \overline {20}^{~2}\cdot 15^{(i)}
\cdot 
\left(\bar 6_1^{(j)}+\hat{\gamma } ~
\frac{\overline \Psi \bar H}{M_P^2}~ 15^{(j)}
\right)\bar H ~,
\label{d54}
\end{equation}

\begin{equation}
\frac{1}{M_P^3}\hat{\Sigma }_i
\cdot \overline {20}^{~3}\cdot 
\left(\bar 6_1^{(i)}+\hat{\sigma } ~
\frac{\overline \Psi \bar H}{M_P^2}~ 15^{(i)}
\right)\bar H^2
\label{d55}
\end{equation}
one can simply verify that in addition to suppression from the CKM matrix 
elements, the ${\cal R}\times  {\cal U}(1)$ symmetry also provides a strong 
suppression.

\section{Neutrino Masses in the Minimal Scheme}

Turning now to neutrino masses, at first glance it seems
from  (\ref {matd}) that the  neutrinos have  large
`Dirac' masses $\nu_L^L N h_u^0$. 
However, we also have 
the term $V\nu^cN$, which is 
crucial for the suppression of the left handed neutrino mass
by the see-saw mechanism \cite{seesaw} .   

The mass matrix which include three families of the appropriate 
matter fields and  is relevant for  neutrino masses is 
$15\times 15$ dimensional and is schematically written as:

\begin{equation}
\begin{array}{ccccc}
 & {\begin{array}{ccccc} 
\nu_L^L & \,\,~~~ \nu_L^{\bar L} & \,\,~~~~\nu_L^l&
\,\,~~~~ \nu^c &\,\,~~~ N
\end{array}}\\ \vspace{2mm}
\hat{M}_{\nu }= \begin{array}{c}
 \nu_L^L \\ \nu_L^{\bar L} \\\nu_L^l
\\\nu^c \\ N \end{array}\!\!\!\!\! &{\left(\begin{array}{ccccc}
\,\, 0 &\,\,~~{\cal C}v  &\,\,~~\hat{\delta }  &\,\,~~0 &\,\,~~\hat{m}' \\ 
\,\, {\cal C}^Tv &\,\,~~0 &\,\,~~{\cal B}V &\,\,~~ 0 &\,\,~~ 0  \\
\,\,\hat{\delta }&\,\,~~{\cal B}^TV &\,\,~~0 &\,\,~~\hat{m}_u^T&\,\,~~0\\
\,\,0&\,\,~~0 &\,\,~~\hat{m}_u &\,\,~~0 &\,\,~~\hat{M} \\ 
\,\,\hat{m}^{'T}&\,\,~~0 &\,\,~~0&\,\,
~~\hat{M}^T & \,\,~~0 \end{array}\right)} 
\end{array}  \!\! ~, ~~~~~
\label{bigmatnu}
\end{equation} 
where, according to  (\ref{matab}) and (\ref{matc}),

\begin{equation}
\nonumber \\
\begin{array}{ccc}
{\cal B}=~~ \\
\end{array}
\hspace{-6mm}\left(
\begin{array}{ccc}
b_{11}\epsilon_X^6 & 0 & 0 \\
0 &b_{22}\epsilon_X^3  & 0 \\
b_{31}\epsilon_X^3 &b_{32}\epsilon_X^2  & b_{33} \end{array}
\right),~~~~
\begin{array}{ccc}
{\cal C}=~~ \\
\end{array}
\hspace{-6mm}\left(
\begin{array}{ccc}
c_{11}\epsilon_X^9 & 0 & c_{13}\epsilon_X^6 \\
c_{21}\epsilon_X^8 & c_{22}\epsilon_X^6 &c_{23}\epsilon_X^5  \\
c_{31}\epsilon_X^7 &0  &c_{33}\epsilon_X^4  \end{array}
\right),~~~~
\label {bc}
\end{equation}
$\hat{m}_u$ is given by  (\ref{matu}), and

\begin{equation}
\nonumber \\
\begin{array}{ccc}
\hat{\delta }=~~ \\
\end{array}
\hspace{-6mm}\left(
\begin{array}{ccc}
0 & 0 & 0 \\
0 & 0 & 0 \\
0 & 0 & -\lambda_t
\end{array}\right)\frac{V}{v^2}(h_u^0)^2,~~~
\begin{array}{ccc}
\hat{m}'=~~ \\
\end{array}
\hspace{-6mm}\left(
\begin{array}{ccc}
0  &(d\hspace{0.5mm}'-d)_{12}\epsilon_X^4  & 0 \\
(d\hspace{0.5mm}'-d)_{21}\epsilon_X^4 & 
(d\hspace{0.5mm}'-d)_{22}\epsilon_X^3  &
(d\hspace{0.5mm}'-d)_{23}\epsilon_X  \\
0 & (d\hspace{0.5mm}'-d)_{32}\epsilon_X^2 & 
-\left( \frac{V}{v} \right)^2\lambda_t \end{array}
\right)\cdot h_u^0,~~
\label{matbloc1}
\end{equation}

\begin{equation}
\begin{array}{ccc}
\hat{M}=~~ \\
\end{array}
\hspace{-6mm}\left(
\begin{array}{ccc}
0 &d\hspace{0.5mm}'_{12}\epsilon_X^4  & 0 \\
d\hspace{0.5mm}'_{21}\epsilon_X^4  &
d\hspace{0.5mm}'_{22}\epsilon_X^3  &
d\hspace{0.5mm}'_{23}\epsilon_X  \\
0& d\hspace{0.5mm}' _{32}\epsilon_X^2 &\lambda_t  \end{array}
\right)\cdot V~.~~~~
\nonumber \\
\label {matbloc2}
\end{equation}
In the second matrix of (\ref{matbloc1}) 
the expression $(d\hspace{0.5mm}'-d)_{ij}$ stands for
$d\hspace{0.5mm}'_{ij}-d_{ij}$ . The $(3,3)$ elements of the 
matrices (\ref{matbloc1}) and (\ref{matbloc2}) 
are obtained 
after integrating out the
$\overline {\nu}^c_{20}$ and $\nu^c_{\overline {20}}$  states 
(which come from
$20$ and $\overline {20}$ respectively.).

Integrating out the states which correspond to the 
(4,5) and (5,4) blocks of the matrix (\ref{bigmatnu}),  
we obtain the reduced mass matrix for the neutrino
masses:

\begin{equation}
\begin{array}{ccc}
 & {\begin{array}{ccc} 
\nu_L^L & \,\,~~~~~ \nu_L^{\bar L} & \,\,~~~~~~\nu \hspace{1mm}'
\end{array}}\\ \vspace{2mm}
\hat{M}'_{\nu }= \begin{array}{c}
\nu_L^L \\ \nu_L^{\bar L} \\\nu \hspace{1mm}' 
 \end{array}\!\!\!\!\! &{\left(\begin{array}{ccc}
\,\, 0 &\,\,~~{\cal C}v &\,\,~~\hat{\Omega }\\ 
\,\, {\cal C}^Tv &\,\,~~0 &\,\,~~{\cal B}V\\
\,\,\hat{\Omega }^T&\,\,~~{\cal B}^TV &\,\,~~0 
\end{array}\right)} 
\end{array}  \!\!~,  ~~~~~
\label{matnu0}
\end{equation} 
where
\begin{equation}
\begin{array}{ccc}
\hat{\Omega }=~~ \\
\end{array}
\hspace{-6mm}
\left(
\begin{array}{ccc}
0 &\epsilon_X^4  & 0 \\
\epsilon_X^4  &\epsilon_X^3  &\epsilon_X  \\
0& \epsilon_X^2 &0  \end{array}
\right)\cdot \frac{(h_u^0)^2}{V}~.~~~~
\nonumber \\
\label {redbloc}
\end{equation}

After the second stage of integration of the (2,3) and (3,2) 
blocks of the matrix in (\ref {matnu0}), the mass matrix 
for the light neutrinos will be:

\begin{equation}
\begin{array}{ccc}
\hat{m}_{\nu }=~~ \\
\end{array}
\hspace{-6mm}\left(
\begin{array}{ccc}
0 &\rho \epsilon_X^2  & 0 \\
\rho \epsilon_X^2  &\sigma \epsilon_X  &1  \\
0& 1 &0  \end{array}
\right)\cdot \frac{\epsilon_X^6}{\epsilon}\frac{(h_u^0)^2}{V},~~~~
\label {matnu1}
\end{equation}
where $\rho $ and $\sigma $ are   couplings of  order unity. 
This matrix is diagonalized through the transformation

\begin{equation}
L_{\nu }^{\dag }\hat{m}_{\nu }L_{\nu }=\hat{m}_{\nu }^{diag}~,~~~~
\begin{array}{ccc}
L_{\nu }=~~ \\
\end{array}
\hspace{-6mm}\left(
\begin{array}{ccc}
1 &\frac{\rho }{\sqrt 2}\epsilon_X^2  &
\frac{\rho }{\sqrt 2} \epsilon_X^2 \\
0 &\frac{1}{\sqrt 2}  &-\frac{1}{\sqrt 2} \\
-\frac{\rho }{2}\epsilon_X^2 &\frac{1}{\sqrt 2} & 
\frac{1}{\sqrt 2}  \end{array}
\right)~,~~~~~
\label {Lnu}
\end{equation}
and the corresponding eigenvalues are 
\begin{equation}
m_{\nu_1}=0~,~~~
m_{\nu_2}\sim m_{\nu_3}\sim
\frac{\epsilon_X^6}{\epsilon}\frac{(h_u^0)^2}{V}
\sim 10^{-5}~{\rm eV}~.
\label{nu}
\end{equation}

With these values of neutrino masses and mixings it is not possible it seems
to explain the atmospheric and solar neutrino deficit, so that some extension
of the model is needed for accommodating the present experimental data
\cite{sup}.

\subsection{Accommodating Atmospheric and Solar Neutrino Data}

To provide an explanation
of the solar and atmospheric neutrino deficit in the framework
of our model we invoke one of the mechanisms described in 
refs. of \cite{mech}.
For our model the most natural scenario is one in which the
atmospheric anomaly is resolved through maximal $\nu_{\mu }-\nu_s$
oscillations, while the solar neutrino puzzle is explained by the small
angle $\nu_e-\nu_{\tau }$ MSW oscillations.

Introducing the sterile neutrino $\nu_s$ and right handed
neutrino $N'$, with the following transformation properties under the 
${\cal R}\times {\cal U}(1)$ symmetry 

$$
{\cal R}({\cal Z})~:~~~~R_{\nu_s}=\frac{1}{2}\alpha ~,~~~~~~~~~~
R_{N'}=\frac{3}{2}\alpha
$$
\beq
{\cal U}(1)~:~~~~R_{\nu_s}=-\frac{185}{14}R~,~~~~
R_{N'}=-\frac{39}{14}R~,
\label{trans}
\eeq
the relevant couplings will be:
$$
W_{N'\nu_s }=\kappa \left(\frac{XZ^2}{M_P^3}15_1+
\frac{X^2}{M_P^2}15_2+\frac{X}{M_P}15_3\right)N'\overline{\Psi }+
$$
\begin{equation}
\left(\frac{Z}{M_P}\right)^4\frac{\bar HHX}{M_P^2}N'^2+
\left(\frac{Z}{M_P}\right)^{16}\nu_s15_2\overline{\Psi }~,
\label{str}
\end{equation}
where $\kappa $ is a dimensionless constant.

Substituting in (\ref{str}) the VEVs of appropriate fields and integrating
out the heavy $N'$ state, the light neutrino mass matrix will have 
the form

\begin{equation}
\begin{array}{cccc}
 & {\begin{array}{cccc}
\hspace{-5mm}~~~\nu_e & \,\,~~~~\nu_{\mu }~~~  &
\,\,~\nu_{\tau }~~& \,\,~\nu_s
\end{array}}\\ \vspace{2mm}
m_{\nu }= \begin{array}{c}
\nu_e \\ \nu_{\mu } \\ \nu_{\tau } \\ \nu_s
 \end{array}\!\!\!\!\! &{\left(\begin{array}{cccc}
\,\,m'\epsilon_X^4  &\,\,~~m'\epsilon_X^3 &
\,\,~m'\epsilon_X^2 &~0
\\
\,\,m'\epsilon_X^3  &\,\,~~m'\epsilon_X^2  & \,\,~m'\epsilon_X&m
 \\
\,\,m'\epsilon_X^2 &\,\,~~m'\epsilon_X &\,\,~m' & 0
\\
\,\,0 &\,\,~~ m &\,\,~0 &~~0
\end{array}\right) }~,
\end{array}  \!\!  ~~~~~
\label{matst}
\end{equation}
where

\beq
m=\epsilon_X^{16}h_u~,~~~~~~~m'=\frac{\kappa^2h_u^2}{M_P\epsilon_X^5}~.
\label{pars}
\eeq
For $\epsilon_X\simeq 0.2-0.22$~, $\kappa \sim 1/3$, from (\ref{matst})
and
(\ref{pars}) we see that $\nu_{\mu }$ form a quasi-degenerate massive
state
with $\nu_s$ ($m_{\nu_2}\simeq m_{\nu_s}\simeq m\sim 1-5$~eV),
while the mass
of the
active neutrino state is 
$m_{\nu_3}\simeq m'\sim 3\cdot 10^{-3}$~eV.
The sterile neutrino state is kept light by the symmetry
${\cal R}\times {\cal U}(1)$\footnote{For obtaining light
sterile states the $R$ symmetry was also used in ref.
\cite{chun}.}.    
For atmospheric and solar neutrino oscillation parameters 
we obtain respectively

$$
\Delta m_{\nu_s2}^2\simeq 3mm'\epsilon_X^2\simeq 
10^{-3}{\rm eV}^2~,
$$
\beq
\sin^2 2\theta_{\mu s}=1-{\cal O}(\epsilon_X^2)~,
\label{atm}
\eeq

$$
\Delta m_{31}^2\simeq m_3^2\simeq 10^{-5}{\rm eV}^2~,
$$
\beq
\sin^2 2\theta_{e\tau }\sim 4\epsilon_X^4\simeq 6\cdot 10^{-3}~,
\label{sol}
\eeq
which are in good agreement with the latest atmospheric \cite{sup}
and solar neutrino data \cite{bac}.
In (\ref{atm}) we have taken $m\simeq 3$~eV. One active neutrino with mass
in this range can contribute roughly $15\%$ to the critical energy density
of the universe.
Models of structure formation with cold and hot dark matter
\cite{hdm} are in good agreement with the observations.

Integration of the heavy $N$, $\nu^c$ states will not change this
picture if we introduce states ${\cal N}_1$, ${\cal N}_2$, with
proper transformation properties, and include couplings 
${\cal N}_115_2\overline{\Psi }$ and ${\cal N}_2\bar 6_1^{(1)}H$
in the theory.
Then it is easy to verify that after decoupling of heaviest states,
the matrix (\ref{matst}) will not be affected and the results 
(\ref{atm} ), (\ref{sol} ) will still hold.

\section{Conclusions}

 Inspired by superstring theories, we have discussed how a realistic `low
energy' phenomenology can arise from a supersymmetric 
$SU(6)\times  U(1)$ gauge 
theory. In order to realize this goal, one must supplement the gauge
symmetry with additional symmetries, and we have provided one example
of this, to wit, the symmetry ${\cal R}\times {\cal U}(1)$, where 
${\cal R}$ denotes a
discrete $R$-
symmetry (it is likely that the continuous ${\cal U}(1)$
symmetry can be replaced
by some discrete symmetry which `effectively' 
behaves as ${\cal U}(1)$). The model
we have presented has several interesting features. The gauge hierarchy
problem is resolved by the pseudo-Goldstone mechanism, 
and the proton lifetime is consistent with observations.
The parameter tan$\beta$ turns out to
be of order unity. Finally, the 
${\cal R}\times {\cal U}(1)$ symmetry also enables us to explain
the observed fermion mass hierarchies and mixings, and it plays an essential
role in providing a `light' sterile neutrino that is 
needed here for a simultaneous explanation
of the solar and atmospheric neutrino puzzles. 
The scenario predicts that
neutrino hot dark matter constitutes roughly $15$-$20\%$ of the critical
energy density.


\vspace{.5cm}
{\bf {Acknowledgements}}

One of us (Q.S.) would like to acknowledge the 
DOE support under grant no. DE-FG02-91ER40626.
We also acknowledge support of NATO Grant CRG-970149.

\end{document}